\documentstyle[psfig]{mn}

\def\spose#1{\hbox to 0pt{#1\hss}}
\def\lta{\mathrel{\spose{\lower 3pt\hbox{$\mathchar"218$}}
     \raise 2.0pt\hbox{$\mathchar"13C$}}}
\def\gta{\mathrel{\spose{\lower 3pt\hbox{$\mathchar"218$}}
     \raise 2.0pt\hbox{$\mathchar"13E$}}}

\title[Distant FR\,I radio galaxies in the HDF]{Distant FR\,I radio galaxies in the Hubble Deep Field:
Implications for the cosmological evolution of radio-loud AGN}

\author[I.A.G. Snellen \& P.N. Best]
{I.A.G. Snellen \& P.N. Best\\ 
 Institute for Astronomy, Royal Observatory, Blackford Hill, 
Edinburgh EH9 3HJ, United Kingdom}

\date{}

\begin{document}
\maketitle

\begin{abstract}
Deep and high resolution radio observations
of the Hubble Deep Field and flanking fields have shown the presence
of two distant edge-darkened FR\,I radio galaxies, allowing for the
first time an 
estimate of their high redshift space density.  If it is assumed
that the space density of FR\,I radio galaxies at z$>$1 is similar to
that found in the local universe, then the chance of finding two FR\,I
radio galaxies at these high radio powers in such a small area of sky
is $<1\%$.  This suggests that these objects were significantly more
abundant at z$>$1 than at present, effectively ruling 
out the possibility that FR\,I radio sources undergo no cosmological
evolution. We suggest that FR\,I and 
FR\,II radio galaxies should not be treated as intrinsically 
distinct classes 
of objects, but that the cosmological evolution is simply 
a function of radio power with  FR\,I and FR\,II radio galaxies of 
similar radio powers undergoing similar cosmological evolutions.
Since low power radio
galaxies have mainly FR\,I morphologies and high power radio galaxies
have mainly FR\,II morphologies, this results in a generally stronger
cosmological evolution for the FR\,IIs than the FR\,Is.
We believe that additional support from the $V/V_{\rm{max}}$ test 
for evolving and non-evolving population of FR\,IIs and FR\,Is 
respectively is irrelevant, since this test is sensitive over very 
different redshift ranges for the two classes.
\end{abstract}

\section{Introduction}

\subsection*{Radio source counts and the cosmological evolution of 
radio galaxies}

Already in the early days of radio astronomy it was realised that 
number counts of radio sources contain important information 
about the distribution of radio sources as a function of redshift.
Early results showed that they were inconsistent with the predictions
of a Steady-state cosmology (Ryle \& Clarke 1961), but that 
they could be explained by evolutionary cosmological models in 
which there were many more powerful radio sources at epochs earlier 
than the present. This could be directly inferred from the slope
of the Log$N-$Log$S$ curve above Jy levels being steeper than 
the $-\frac{3}{2}$ power law which the static, uniform, 
Eucleidian Universe must show. At sub-Jy levels the slope changes 
gradually to a law flatter than the $-\frac{3}{2}$ law as it must
to escape the radio equivalent of Olber's Paradox. Longair (1966)
showed that the strong evolution must have been confined to 
the most powerful radio sources, because 
the observed range in power for
radio sources was more than 5 orders of magnitude, 
while the turnover in the radio source counts covered less than 
2 orders of magnitude.

The initial steep slope and its turnover were subsequently 
accounted for by two distinct populations of radio sources.
One population consisted of high radio luminosity sources 
undergoing strong cosmological
evolution (e.g. with comoving number densities about $2-3$ orders 
of magnitude higher at large redshifts), 
the other population was of low luminosity sources 
undergoing no or only little 
evolution with cosmological epoch. Wall (1980) suggested that 
these two populations of non-evolving and strongly 
evolving radio sources corresponded to the 
Fanaroff \& Riley (1974) Class I and II galaxies respectively.
This idea has been further developed by Jackson and Wall (1999),
who use the source counts at several frequencies and redshift
information of the 3C sample, to link the FR\,I and FR\,II radio 
sources as the unbeamed parent populations of BL Lac objects and flat
spectrum quasars respectively. 
Willott et al. (2000) also defined a dual-population scheme, 
but slightly different from that of Jackson \& Wall (1999).
In their scheme, the low luminosity population is associated with
radio galaxies with weak emission lines (both FR\,Is and FR\,IIs),
and the high luminosity population is associated with radio galaxies
and quasars with strong emission lines (almost all FR\,IIs). 
In contrast, Dunlop \& Peacock (1990) do not make a distinction
between FR\,I and FR\,IIs, in their analysis of the statistics 
of steep spectrum radio sources, but allow for a luminosity
dependent cosmological evolution. 

\subsection*{Fanaroff \& Riley Class I and II radio sources}

Fanaroff \& Riley (1974) showed that there is a distinct 
morphological difference between radio sources of high and low
luminosity. The more powerful sources have their regions of 
highest surface brightness at the ends of a double lobed
structure (FR II $-$ edge brightened), while the lower power objects 
show a variety of forms in which the highest brightness 
occurs near their centres, 
excluding their cores (FR I $-$ edge-darkened).
Fanaroff \& Riley (1974) found a sharp division between the two 
classes at 
$P_{\rm{178 MHz}} \sim 2\times10^{25}$ W Hz$^{-1}$sr$^{-1}$.

Early optical work indicated that there is also a difference between
the host galaxies of FR\,I and FR\,II type radio galaxies.  Owen and
Laing (1989) found that FR\,II sources reside in normal giant
elliptical galaxies with absolute magnitudes near $M_{\rm{*}}$ of the
Schechter luminosity function, considerably fainter than first-ranked
galaxies in rich clusters, while FR\,I sources reside in galaxies
which can generally be described as D or cD galaxies.  However, more
recently it has been shown by Ledlow and Owen (1996) that this result
was caused by the combination of a sample selection effect and a
strong positive correlation between the FR\,I/II radio luminosity
cutoff and the absolute magnitude of the host-galaxy, with an increase
of the transition luminosity from
$L_{\rm{1.4GHz}}$$\approx$$10^{24}$ W/Hz at M$_R$=$-$21, to
$L_{\rm{1.4GHz}}$$\approx$$10^{26}$ W/Hz at M$_R$=$-$24. 
Due to the small
range of radio-luminosity in the Owen and Laing sample, this
correlation resulted in the FR\,IIs generally residing in lower
luminosity galaxies than the FR\,Is.  The Ledlow \& Owen result
suggests that both FR\,I and FR\,II radio sources live in similar
environments, but that the properties of the host galaxies may
strongly influence the morphological appearance of the radio sources,
at least for those near the division of the two classes.

Zirbel \& Baum (1995) found that for the same total radio power, 
FR\,II galaxies produce 5$-$30 times more emission-line luminosity
than FR\,I galaxies. In this light, Baum, Zirbel \& O'Dea (1995)
put forth the possibility that the FR dichotomy 
is due to qualitative differences in the structural properties
of the central engines in these two types of sources, like
the accretion rate and/or the spin of the central black hole. 
In contrast, Gopal-Krishna \& Wiita (2000) point out a 
class of double radio sources in which the two
lobes exhibit clearly different FR morphologies. Although these objects 
are rare, their existence supports explanations for
the FR dychotomy based upon jet interaction with the external medium,
and appears quite difficult to reconcile with the class of
explanations that posit fundamental differences in the central engine.
 
\subsection*{Radio Observations of the Hubble Deep field}

\begin{table*}
\begin{tabular}{lccrrl}
Telescope & Frequency & Resolution & Flux limit & Number of & Reference\\
          &  (GHz)    & (arcsec)   & ($\mu$Jy)  &  Sources  &        \\

VLA A,CnB, C, DnC \& D & 8.4  & 0.3$-$10&  9 & 29 & Richards et
al. (1998)\\
VLA + MERLIN& 1.4  & 0.2     & 40 & 87 & Muxlow et
al. (2001)\\
WSRT        & 1.4  & 15      & 40 & 85 & Garrett et
al. (2000)\\
EVN         & 1.4  & 0.025   & 210  & 3 & Garrett et
al. (2001)\\
\end{tabular}
\caption{\label{radio}Radio observations of the Hubble Deep Field
and Flanking Fields. The columns give the telescope used,
the observing frequency, resolution, flux density limit, number of 
detected sources, and the reference to the observation.}
\end{table*}

Due to observational limitations, no direct measurements
of the high-z space density of FR\,I radio galaxies yet exist. 
The Hubble Deep Field (HDF; Williams et al. 1996) and
the surrounding Hubble Flanking Field (HFF) is the best 
studied area of sky to date. The random region is unbiased,
although it was chosen for  
the absence of any bright objects at any wavelength. 
Recently, several groups have observed the HDF in the 
radio regime, using the VLA, MERLIN, WSRT and the EVN
(see table \ref{radio}). This allows a first estimate of the
high redshift space density of FR\,I radio sources, which is 
discussed in this paper.
 
Richards et al. (1998) have observed the HDF at 8.4 GHz using
the VLA in A, CnB, C, DnC and D configurations, corresponding to 
angular resolutions ranging from 0.3$''$ to 10$''$. They detected 29 
radio sources in a complete sample within 4.6$'$ of the HDF centre
and above a flux density limit of 9 $\mu$Jy, of which 7 are located
in the HDF itself. 
Muxlow et al. (1999, 2001) have observed the HDF at 1.4 GHz,
using the VLA for 42 hours and MERLIN for 18 days.
They have detected a complete sample of 87 sources 
in a $10'\times10'$ region with flux 
densities above $40\mu$Jy. These have all been imaged with the 
MERLIN+VLA combination to produce images with 0.2, 0.3, and 0.5$''$
resolution with an rms noise-level of $3.3 \mu$Jy. 
In addition, Garrett et al. (2000) have observed the HDF with the 
WSRT at 1.4 GHz, resulting in an angular resolution of 15$''$ and a 
noise-level of 8 $\mu$Jy.
 They detect 85 regions of radio emission ($>5 \sigma$) in a 
$10'\times10'$ field centered on the HDF, with 4 sources 
not previously detected at 1.4 GHz.  
Garrett et al. (2001) have taken deep European VLBI Network (EVN) 
observations of the HDF, with a resolution of 25 milliarcseconds. 
This has resulted in the detection of 3 sources above a flux density
level of 210 $\mu$Jy.

The deep optical/radio observations of the HDF and HFF
 indicate that  $\sim$ 60\% of the faint sub-mJy and 
$\mu$Jy radio
sources are identified with star forming galaxies at moderate 
redshifts ($z \sim 0.2-1$), often showing morphologically peculiar,
interacting/merging galaxies with blue colours and HII-like
emission spectra (Richards et al. 1998). Another 20\% of the sources
are as yet unidentified in the optical 
down to I=25.5 in the HFFs and I=28.5 in the HDF. These faint 
systems may be distant galaxies obscured by dust.
The remaining 20\% of the faint radio population 
seem to be identified with relatively low luminosity AGN.

This paper concentrates on
the two brightest radio sources at 1.4 GHz, 
both exhibiting a clear FR\,I radio morphology. We will discuss
the implications of their presence in the HDF and HFFs 
for the cosmological evolution of these objects.
Throughout this paper we assume a cosmology with H$_0$=50 km
sec$^{-1}$ Mpc$^{-1}$ and q$_0$=0.5, for consistency with
previous studies.

\begin{figure*}
\hbox{
\psfig{figure=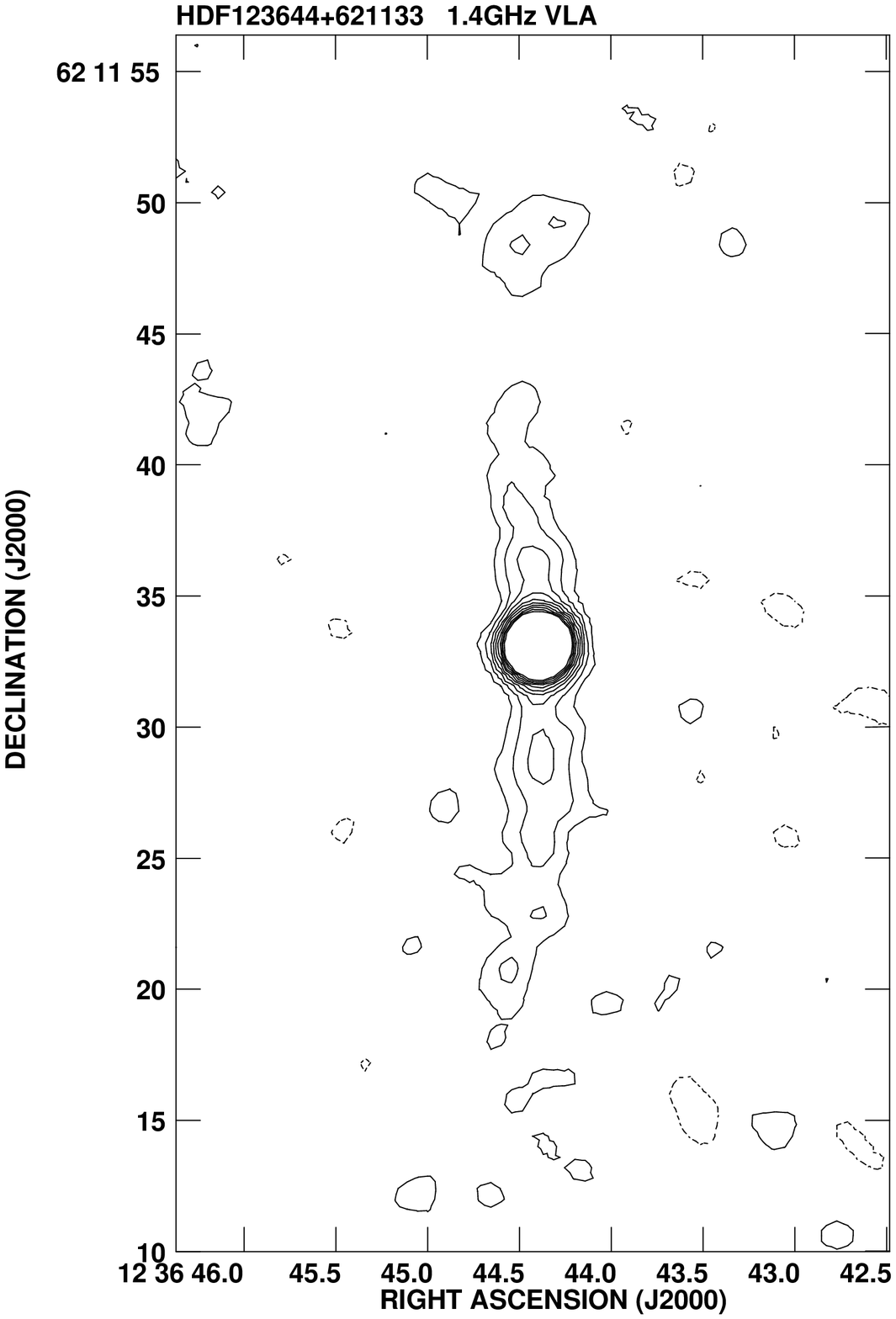,width=7.5cm}
\psfig{figure=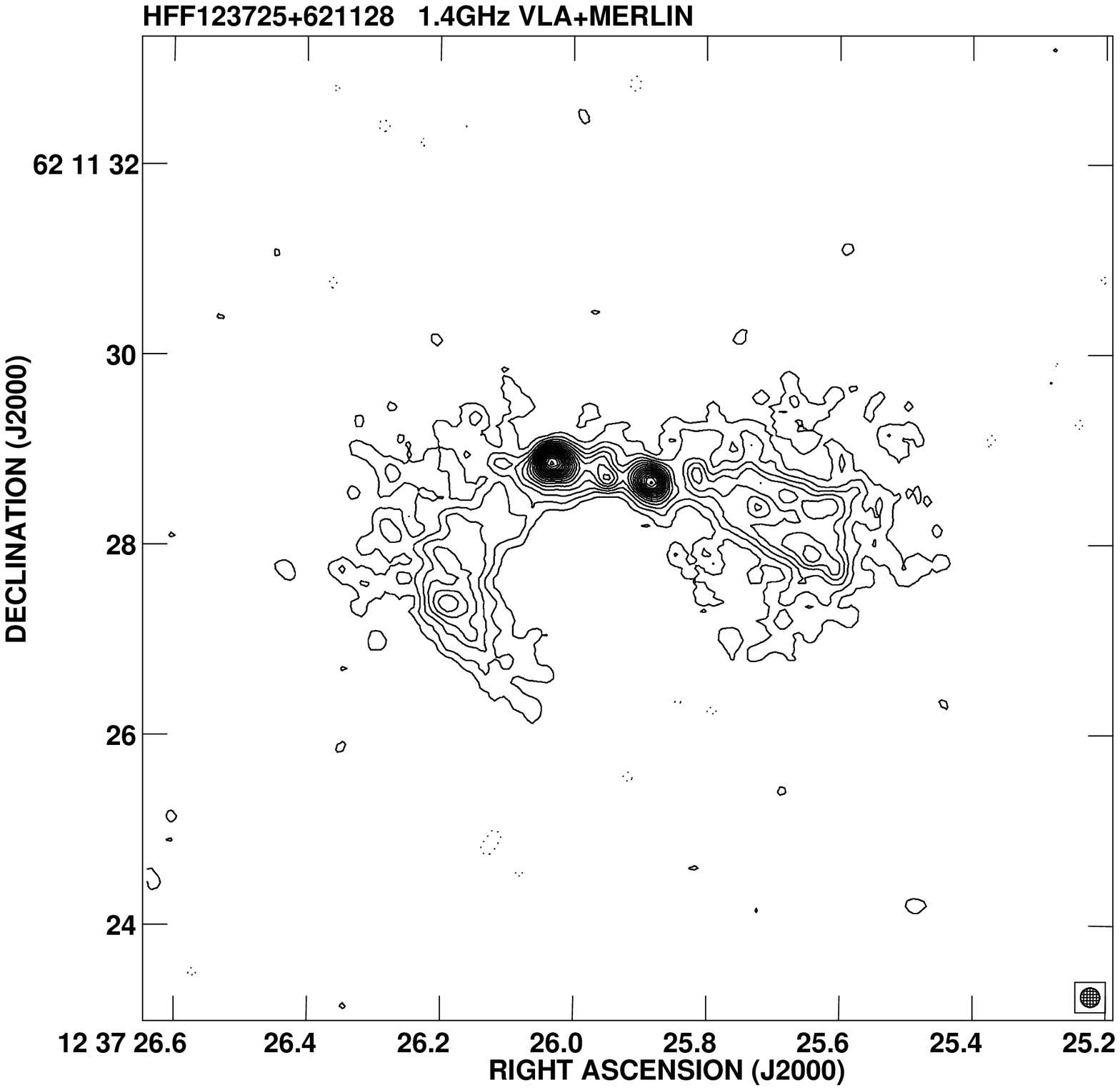,width=10cm,rheight=10.3cm}
}
\caption{\label{vla} The 1.4 GHz image of the FR\,I radio
         sources HDF123644+621133 from VLA-data (left) and of
HFF123725+621128 from VLA+MERLIN data (right), kindly provided
by T. Muxlow (Muxlow et al., in preparation 2001).}
\end{figure*}

\section{FRI radio galaxies in the HDF and HFFs}

\subsection*{HDF123644+621133}

The radio source HDF123644+621133 is the brightest object
in the HDF at 8.4 GHz. It has an 8.4 GHz flux density
of 477 $\mu$Jy. Muxlow et al. (2001) give a total 1.4 GHz
flux density of 1.2 mJy using VLA+MERLIN data, 
while Garrett et al. (2000) report a total flux density of 
1.6 mJy using the WSRT. This higher flux density is 
probably a result of the lower resolution of the 
WSRT observations, containing a larger proportion of the 
total source flux. The radio source consists of
an unresolved core with a relatively flat spectrum 
($\alpha_{\rm{1.4-8.4GHz}}=-0.3\pm0.2$, with 
$f(\nu)\propto \nu^\alpha$), which
is surrounded by steep spectrum 
$\alpha_{\rm{1.4-8.4GHz}}=-1.2\pm0.2$ 
emission oriented north-south and extending about 30$''$ 
(Fig \ref{vla}). The radio source shows a prototypical FR\,I 
morphology. It is optically identified with a bright
$R_{\rm{AB}}=20.5$ elliptical galaxy at z=1.013 
(Richards et al. 1998).
Assuming the flux densities and spectral indices of the core 
and extended structure given above, HDF123644+621133 has 
an observed flux density at 178 MHz rest wavelength 
(88 MHz observed) of $S_{\rm{178 MHz}}=24$ mJy.
This corresponds to a total radio power of 
$P_{\rm{178 MHz}}=10^{24.8}$ W Hz$^{-1}$sr$^{-1}$.

\subsection*{HFF123725+621128}

The radio source HFF123725+621128 is the brightest radio source in 
the HDF and HFFs at 1.4 GHz,
 with a total flux density of $\sim 6$ mJy.
The high resolution MERLIN+VLA image, crucial for a morphological
classification, shows it to be 5$''$ in extent with two high
brightness regions (peak brightness of 200 $\mu$Jy/beam) close to 
its centre, making it an FR\,I radio source (Fig. \ref{vla}). The 
two diffuse extended lobes at the east end west side are both 
slightly bent towards the south giving the object the appearance 
of a wide angle tail (WAT), which at low redshift are mostly found 
in galaxy-cluster environments. Richards et al. (1998) quote
a spectral index for this source of 
$\alpha_{\rm{1.4-8.4GHz}}=-1.0\pm0.1$. Its detection 
in the WENSS survey at 325 MHz (Rengelink et al. 1997) of 
25 mJy is consistent with this.

The radio source is optically identified with a compact, possibly
elliptical galaxy located between the two regions of high brightness
(Richards et al. 1998). It has an $R_{\rm{AB}}$ magnitude of 24.1,
a $G_{\rm{AB}}$ magnitude of $\sim$26.5, 
and an H+K magnitude of 18.7, which corresponds to K$\approx$19.0.
To date no redshift has been measured. 
However, its faint optical magnitude and its red optical to 
near-infrared colours makes it highly unlikely it is at $z<1$.
Comparison with the K-z 
diagram for powerful radio galaxies (eg. Lilly \& Longair 1984), 
also makes it most likely that this object is located at $z>1$.
If we assume a lower limit in redshift of z=1, its 
rest-frame 178 MHz power must be higher than  
$P_{\rm{178 MHz}}\gta 10^{25.3}$ W Hz$^{-1}$sr$^{-1}$, 
making it at least a few times more luminous than HDF123644+621133. 
If it is located at z=1.5, it would have 
$P_{\rm{178 MHz}}\approx 10^{25.7}$ W Hz$^{-1}$sr$^{-1}$. This is
such a high luminosity that it would be comparable to 
the most powerful FR\,Is  found in the local universe, 
ie. those near the FR\,I/II division with the 
 optically most luminous host galaxies (Ledlow \& Owen, 1996).  

\subsection*{Other FR\,I radio sources in the HDF?}

To identify a radio source at high redshift as one with
an FR\,I morphology, the radio observations have to be of
both sufficient resolution (eg. using MERLIN) and of sufficient
sensitivity to detect the low level extended emission. 
The edge-darkened emission, characteristic for an FR\,I radio 
source, is in the case of HDF123644+621133 and HFF123725+621128
only detected a  few $\sigma$ level,
while these sources are two of the brightest objects 
in the HDF and HFFs.
Muxlow et al. (1999) have identified 14, much fainter sources 
which appear to have compact two sided emission. 
For a significant fraction of those, fainter emission on larger
extended scales may have been missed, hampering their classification 
as an FR\,I. We therefore believe that using the existing 
VLA and MERLIN data, only
sources down to mJy levels can reliably be classified as FR\,Is.

\section{The cosmological evolution of FR\,I radio sources}

\subsection*{The local space density of FR\,I radio sources}

We first want to assess the question of how many luminous FR\,I radio 
galaxies one would 
expect in an area of sky as large as the HDF + HFF for a non-evolving
cosmological evolution scenario. As a luminosity cut-off we 
take the luminosity of HDF123644+621133, which is the 
weaker of the two, independent of the exact redshift of 
HFF123725+621128. 

Firstly, the local space density
of FR\,I radio sources has to be determined.
We used the complete subsample of 3CR as defined by Laing et al. 
(1983) for this purpose. It identifies 30 FR\,I radio galaxies
in an area of sky with $\delta >10^\circ$ and $|b|>10^\circ$
(4.24 steradians). 
Using the 3CR flux density cut-off of 10.9 Jy at 178 MHz and an 
average spectral index of $\alpha=-0.75$, a radio galaxy
with $P_{\rm{178 MHz}}=10^{24.8}$ W Hz$^{-1}$sr$^{-1}$
can be seen out to z=0.042.
Four FR\,I galaxies are found 
with $P_{\rm{178 MHz}}>10^{24.8}$ W Hz$^{-1}$sr$^{-1}$
in the comoving volume of 
$2.2\times10^7$Mpc$^3$ ($z<0.042$), which
corresponds to a local space density of 170 FR\,Is per Gpc$^3$.
Note that we obtain a very similar value of 200 
FR\,Is per Gpc$^3$ 
brighter than $P_{\rm{178 MHz}}>10^{24.8}$ W Hz$^{-1}$sr$^{-1}$, 
using the local luminosity function of Dunlop \& Peacock 
(1990; DP90) of steep spectrum radio sources, 
assuming that all sources with 
$P_{\rm{2.7GHz}}>10^{24.2}$ W Hz$^{-1}$sr$^{-1}$
($\equiv P_{\rm{178MHz}}>25.3$ W Hz$^{-1}$sr$^{-1}$) 
are FR\,II, and all the fainter sources are FR\,Is.
We will use the latter value in our calculations.

\subsection*{Implication for a non-evolution scenario for FR\,I radio
galaxies}

Hence, what is the chance of finding 2 FR\,I radio sources 
 in an area as 
large as the HDF and HFF for a non-evolving population?
The area of sky surveyed by the VLA and MERLIN covers 
$10' \times 10'$. 
The main parameter involved is the 
maximum redshift, $z_{\rm{max}}$, at which the radio sources 
still would have been detected and classified as FR\,Is.
 To estimate this is a difficult task, 
not the least due to the uncertain redshift of one of the objects.
 We believe that the limiting factor for $z_{\rm{max}}$ for both
sources is the low surface brightness level of the extended
emission in their lobes, which is proportional to $(1+z)^{-4}$,
assuming a spectral index in the lobes of $\alpha=-1.0$.

\begin{table*}
\caption{\label{calc} Chance of finding 2 FR\,Is in the HDF and HFF 
with $P_{\rm{178 MHz}}>24.8$ W Hz$^{-1}$sr$^{-1}$ for 
no evolution, and the pure luminosity and luminosity/density
evolution models of Dunlop \& Peacock (1990), 
using different redshift cut-offs.}
\begin{tabular}{lllrllrllr}\hline
          &&\multicolumn{2}{l}{No Evolution}  
&\multicolumn{3}{l}{Pure Lum. Evolution}
&\multicolumn{3}{l}{Lum/Dens Evolution}\\
z range & Volume    & Exp. &$\rm{P}_{>2}$& Density   & Exp. &$\rm{P}_{>2}$&Density   & Exp. &$\rm{P}_{>2}$\\
          & (Mpc$^3$) & nr   &      & enhance. & nr   &      & enhance. & nr   &      \\
$<$1.5 & $2.4\times10^5$ &0.048 &0.1\% &  4.8      &0.23  & 2\%& 4.1&0.20&2\%\\
$<$2.0 & $3.7\times10^5$ &0.074 &0.3\% &  6.4      &0.47  & 8\%& 5.5&0.41&6\%\\
$<$2.5 & $4.9\times10^5$ &0.098 &0.4\% &  7.6      &0.74  &17\%& 6.6&0.65&14\%\\
$<$3.0 & $6.1\times10^5$ &0.122 &0.7\% &  8.2      &1.00  &26\%&
7.0&0.85&21\%\\ \hline
\end{tabular}
\end{table*}

In both objects, the extended emission is detected at about a 
10 $\sigma$ level, implying that HDF123644+621133
could have been detected out to z$\approx$1.7 at $>3\sigma$.
We do not know the redshift of HFF123725+621128, but as we 
argued above, we believe it will probably be in the range 
1$<$z$<$1.5$\sim$2.
If it were at z=1, it would have a $z_{\rm{max}}$ of $\sim$1.7.
If it were at z=1.5 or z=2.0, it would have a $z_{\rm{max}}$ of 
$\sim$2.4 or $\sim$3.0. We therefore do the calculation 
for $z_{\rm{max}}=$ 1.5, 2.0, 2.5 and 3.0.
The data are given in table \ref{calc}, with the $z_{\rm{max}}$ in 
column 1, the comoving volume densities in column 2, 
the expected number of FR\,Is in the HDF \& HFF
with $P_{\rm{178 MHz}}>24.8$ W Hz$^{-1}$sr$^{-1}$ in column 3.
The chance, $P_2$, to find {\it two or more} objects in this volume, 
if the expected number is $p$ is, 
\begin{equation}
P_{>2}=e^{-p}(e^p-1-p)
\end{equation}
which is derived using a Poisson distribution.
This is given in column 4 of table \ref{calc}.
It indicates that the presence of two FR\,I radio galaxies in the 
HDF and HFFs is inconsistent ($<1\%$) with a no-evolution scenario 
for FR\,Is, and that their comoving space density at 
$z=1-2$ was about an order of magnitude higher than at present time. 

\subsection*{Constraints on other cosmological evolution models}

Now that we have determined that it is unlikely that 
FR\,I radio sources undergo no cosmological evolution from z=0
to z=1, we want to investigate whether FR\,I and FR\,II radio 
sources have to be treated as to distinct populations
of objects. In this section we will compare the data with the 
evolutionary models of DP90.
DP90 do not treat the FR\,Is and FR\,IIs separately;
instead they  adopt a 
luminosity-dependent cosmological evolution for the 
total population of steep spectrum sources, derived
from source counts and redshift information.
We will consider two of their models: the pure luminosity evolution
model and the luminosity/density evolution model.

In both cases, the low redshift luminosity function (LF) is fitted
using two powerlaw slopes, with the LF being flatter below the break
luminosity and steeper above.
In the pure luminosity evolution model, the overall shape of this
LF does not change with cosmological epoch, 
only the normalisation in luminosity. 
Since the slope of the 
LF is flatter at low luminosities, it results in less cosmological
evolution for the weaker objects.
Since we only have to consider objects below the the break 
luminosity, where the slope of the luminosity function 
is $\alpha=-0.69$,
at a specific redshift the density is enhanced by a factor
$10^{0.69}*f$. The parameter $f$ is the amount of luminosity 
evolution, which is determined by DP90 to be $f=1.26z-0.26z^2$.
For the redshift ranges involved, the local space density
of FR\,Is would on average be enhanced by a factor of $5-8$, 
which corresponds to $P_{>2}=2-26\%$. The data are given in table
\ref{calc}, with column 5 giving the mean density enhancement,
column 6 giving the expected number of FR\,Is in the HDF \& HFF, 
and column 7 giving the chance of finding two or more FR\,Is in
this area, for the pure luminosity evolution model of DP90.

The luminosity/density evolution model of DP90 is based on their
pure luminosity evolution model, but modified to allow also 
the density normalisation to vary with z.
In a similar fashion as above, but with slightly different
parameters, this model would result in an average space density
enhancement of a factor 4 to 7, which implies a 
chance of finding two or more FR\,Is in
this area of 2$-$21\%, for the luminosity/density evolution model 
of DP90. This is given in columns 8$-$10 of table \ref{calc}.

Therefore, in contrast to the non-evolution models, the DP90
models, which do not treat FR\,I and FR\,II
as distinct populations, predict surface densities of FR\,I
radio sources which are only slightly lower than  
found in the HDF and HFF.

\begin{figure}
\psfig{figure=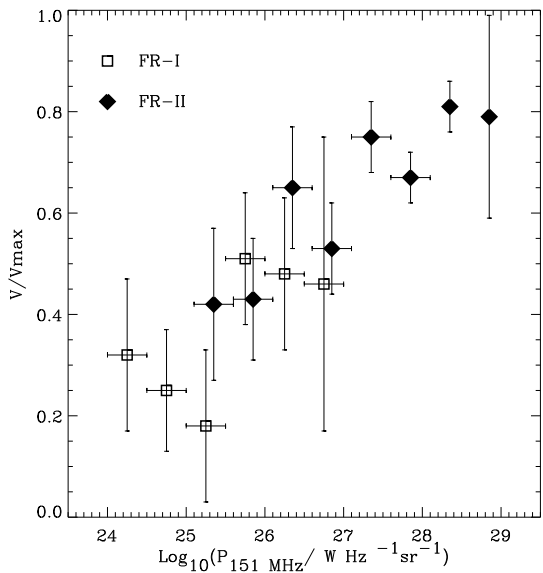,width=8cm}
\caption{\label{vvmax} Average V/V$_{\rm{max}}$ values for 
FR\,Is (squares) and FR\,IIs (diamonds), for bins in 
$\Delta \rm{log}_{10}(P_{\rm{151MHz})} = 0.5$. These values are 
taken from Jackson and Wall (1999). Note that no distinction can
be made between FRIs and FRIIs.}
\end{figure}

\section{Discussion and conclusions} 
\subsection*{A single population scheme?}

The deep optical/radio observations of the HDF and HFF
have allowed for the first time an estimate of the space density
of FR\,I radio sources at high redshifts. The two FR\,I radio 
sources present in this area of sky indicate that it is 
unlikely that FR\,I radio sources undergo no cosmological evolution.
We believe that this may undermine the reason for having a ``dual
population'' scheme, in which the FR\,I and FR\,II radio sources
are treated as separate classes of objects, with little or 
no evolution and strong cosmological evolution respectively
(eg. Jackson \& Wall 1999). 
We want to advocate a ``single population'' scheme, in which 
the cosmological evolution is a function of radio power, but not
dependent on FR\,class. Since powerful radio sources undergo 
a stronger evolution than the less powerful ones, and FR\,Is
are mainly radio sources of low power, FR\,Is undergo less 
evolution than FR\,IIs. However, the 
populations of FR\,I and FR\,IIs of 
similar radio power undergo similar cosmological evolution.
Indeed, the evolution models of Dunlop \& Peacock (1990),
which make no distinction between FR\,classes, do fit 
number-count statistics and redshift distributions well. 
In addition they predict number densities which are 
comparable to that found in the HDF and HFF. 

Basically, it is the coincidence of two observations which 
have led to a dual population scheme: 1) the V/V$_{\rm{max}}$ test 
showing that high power radio sources undergo strong space density 
enhancements and low power sources do not, and 2) the 
morphological dichotomy between sources of high and low radio power.
However, the V/V$_{\rm{max}}$ tests for FR\,I and 
FR\,II are sensitive over very different redshift ranges. 
Jackson and Wall (1999) show in their figures 2 and 3 the 
mean V/V$_{\rm{max}}$ values as a function of radio power for 
FR\,Is and FR\,IIs respectively, indicating an increase
in V/V$_{\rm{max}}$ as function of radio power. 
We reproduce these values in figure \ref{vvmax}, using different
symbols for the two FR\,classes. As can be seen, no distinction 
can be made between the mean V/V$_{\rm{max}}$ values 
of FR\,I and FR\,II sources at a particular radio power. It is 
due to the fact that FR\,Is are of lower radio power, and therefore
only found at the lowest redshifts, that they show a
lower average V/V$_{\rm{max}}$. We therefore believe that the
results of the V/V$_{\rm{max}}$ test
actually do not make a dual population scheme necessary, and
perhaps even argue against one.

Willott et al. (2001) also introduced a dual population scheme
in their analysis of complete flux density limited samples
of 3C, 6C and 7C, which provided an unprecedented coverage of 
the redshift-luminosity plane. Their division between 
low power radio sources with weak emission lines and 
high power radio sources with strong emission lines 
(not by FR\,class), results in a space density enhancement 
of about an order of magnitude for the weak objects out to z=1, and 
about three orders of magnitude for the powerful objects out to z=2.
Using the current data it is difficult
to make a distinction between this particular scheme and a single 
population scheme as proposed here: although the 
two schemes have conceptually different viewpoints, they 
result quantitatively in a very similar cosmological evolution
for the FR\,I galaxies. Note however, that the evolution as proposed
by Willott et al. (2001) results in a peculiar luminosity
function at high redshift, with a prominent `hump' at the location of
the break-luminosity at low redshift.
  
Treating the population of powerful radio-loud AGN as a 
single class of object would have many benefits, since in this way
the two FR classes are closely linked. A popular paradigm is that
the population of radio-loud AGN come with a range of jet outputs, 
of which the more powerful
may be strong enough to maintain their integrity  until they impact
on the intergalactic medium (IGM) in a shock. This results in an
FR\,II. However, if the jet is too weak, 
it may dissipate its energy by entraining IGM material, 
resulting in a more turbulent FR\,I. This may also explain the 
dependence of the FR\,I/FR\,II radio power divide on the luminosity
of the host galaxy: in more luminous galaxies, which
reside in denser environments or which have denser ISM, 
only jets of higher power can keep their integrity.

If the dual population scheme were to be correct, and 
FR\,Is and FR\,IIs undergo different cosmological evolutions, 
it would imply fundamental differences for these two classes
of object, like the properties of their central engines such as
black-hole spin or jet composition. 
As has been argued by Gopal-Krishna \& Wiita (2000),
the existence of radio sources, in which the two lobes exhibit 
clearly different FR morphologies, are difficult to reconcile with 
such models, but support explanations for the FR dichotomy based upon
jet interaction with the medium external to the central engine. 

\section*{Acknowledgements}

We like to thank Tom Muxlow for providing the VLA-MERLIN images.  
We thank Jim Dunlop and the referee, Julia Riley,
 for a careful reading of the manuscript and helpful comments.
{}
\end{document}